\pdfoutput=1
\documentclass[12pt,preprint]{aastex}
\usepackage[margin=0.75in]{geometry}
\usepackage{graphicx}
\usepackage{natbib}
\usepackage{aas_macros}
\usepackage[section] {placeins}
\usepackage{subfigure}
\usepackage{float}
\usepackage{color}

\newcommand{\kms}{\, {\rm km\, s}^{-1}}

\newcommand{\mpc}{\, {\rm Mpc}}
\newcommand{\kpc}{\, {\rm kpc}}
\newcommand{\hmpc}{\, h^{-1} \mpc}

\newcommand{\hi}{\mbox{H\,{\scriptsize I}\ }}
\newcommand{\heii}{\mbox{He\,{\scriptsize II}\ }}

\def\h2{${\rm\,H_2}$}

\makeatletter 

\def\hi{\noindent \hangindent=2.5em}

\def\kpc{{\rm\,kpc}}

\def\mpc{{\rm\,Mpc}}

\def\kms{{\rm\,km/s}}

\def\vol#1  {{{#1}{\rm,}\ }}

\def\hi{{H\thinspace I}\ }

\def\heii{{He\thinspace II}\ }

\newcount\refno
\newcount\rfno
\def\eq{$^{\the\refno\ }$\advance\refno by 1}
\def\ad{\advance\rfno by 1}

\def\clock{\count0=\time \divide\count0 by 60
     \count1=\count0 \multiply\count1 by -60 \advance\count1 by \time
     \number\count0:\ifnum\count1<10{0\number\count1}\else\number\count1\fi}

\def\myputfigure#1#2#3#4#5%
{\hskip0.03\textwidth\vskip#5pt
\makebox[0pt]{\hskip#2in
\includegraphics[width=#3\textwidth]{#1}}\vskip#4pt\hfill}

\newcount\refno
\newcount\rfno
\def\eq{$^{\the\refno\ }$\advance\refno by 1}
\def\ad{\advance\rfno by 1}

\definecolor{burntorange}{rgb}{1,0.4,0.2}

\begin{document}

\title{Gaussian Random Field: Physical Origin of Sersic Profiles}

\author{
Renyue Cen$^{1}$
} 

\footnotetext[1]{Princeton University Observatory, Princeton, NJ 08544;
 cen@astro.princeton.edu}

\begin{abstract} 

While the Sersic profile family provide adequate fits for the surface brightness profiles of observed galaxies,
the physical origin is unknown.
We show that, if the cosmological density field are seeded by random gaussian fluctuations,
as in the standard cold dark matter model, 
galaxies with steep central profiles have simultaneously extended envelopes of shallow profiles in the outskirts,
whereas galaxies with shallow central profiles are accompanied by steep density profiles in the outskirts.
These properties are in accord with those of the Sersic profile family.
Moreover, galaxies with steep central profiles form their central regions in smaller denser subunits 
that possibly merge subsequently, which naturally leads to formation of bulges.
In contrast, galaxies with shallow central profiles form their central regions in a coherent fashion without significant substructure, 
a necessary condition for disk galaxy formation.
Thus, the scenario is self-consistent with respect to the 
correlation between observed galaxy morphology and Sersic index.
We predict further that clusters of galaxies should display a similar trend, which 
should be verifiable observationally.

\end{abstract}

\section{Introduction}

The process of galaxy formation has likely imprinted useful information in the stellar structures.
A great amount of effort has been invested in characterizing detailed stellar structures of galaxies of all types,
dating back to \citet[][]{1911Plummer} for globular clusters and \citet[][]{1913Reynolds} for the Andromeda,
and if one is so inclined, to \citet[][]{1755Kant} who might be the first contemplating the shape of the Milky Way 
and island universes.
In modern times, among the best known examples, the \citet[][]{1948deVaucouleurs} law - surface brightness $I(R)\propto e^{-kR^{1/4}}$ (where $R$ is radius and $k$ 
a normalization constant) - 
describes giant elliptical galaxies well, whereas the \citet[][]{1962King} law appears to provide better fits for fainter elliptical galaxies;
disk galaxies are in most cases described by the exponential disk model \citep[][]{1971Hodge}: $I(R)\propto e^{-kR}$. 
The major advantage of \citet[][]{1968Sersic} profile family - $I(R)\propto e^{-kR^{1/n}}$ - 
is that they provide an encompassing set of profiles with $n$ from less than 1 to as large as 10,
including the exponential disk ($n=1$) and de Vaucouleurs ($n=4$) model.

Even at the age of sophisticated hydrodynamic simulations,
the physical origin of the Sersic profile family that have well described all galaxies remains enigmatic.
This author is of the opinion that the nature of galaxy formation process in the context of modern 
cosmological structure formation model is perhaps too complex to warrant any possibility of 
analytic fits to be accurate beyond the zero-th order.
While efforts to characterize deviations from or additions to the standard fits
are not only necessary but also very important to account for rich galaxy data \citep[e.g.,][]{1995Lauer},
it would also seem beneficial to construe the basic trend 
displayed by the wide applicability of the Sersic profile family 
to enhance our physical understanding of the galaxy formation process.

In this {\it Letter} we provide a basic physical understanding of the Sersic profile family
in the context of the standard cosmological model with gaussian random density field.
Our simple analysis provides, for the first time,
a self-consistent physical origin for the Sersic profile family.
This also opens up the possibilities to explore the physical links to 
other properties of galaxies, since, for example, it comes natural and apparently inevitable
that the steep profiled galaxies have a much higher fraction of substructures that form early and interactions/mergers
among them would lead to formation of elliptical galaxies,
enabling a self-consistent picture.
This study is the fifth paper in the series ``On the Origin of the Hubble Sequence".

\section{Gaussian Random Field and Sersic Profiles}

The standard cosmological constant dominated cold dark matter 
cosmological model
has a number of distinct features.
One of the most important is that the initial density fluctuations are gaussian and random.
As a result, the statistical properties are fully determined by a vector quantity,
namely, the linear power spectrum of the density fluctuations, $P_k$,
which is well determined by observations from the microwave experiments and others \citep[e.g.,][]{2011Komatsu}.
Observational evidence is that allowed deviations from gaussianity are at the level of $10^{-3}$ and less in the linear regime \citep[][]{2013Ade}.

\begin{figure}[ht]
\centering
\vskip -0.0cm
\resizebox{6.0in}{!}{\includegraphics[angle=0]{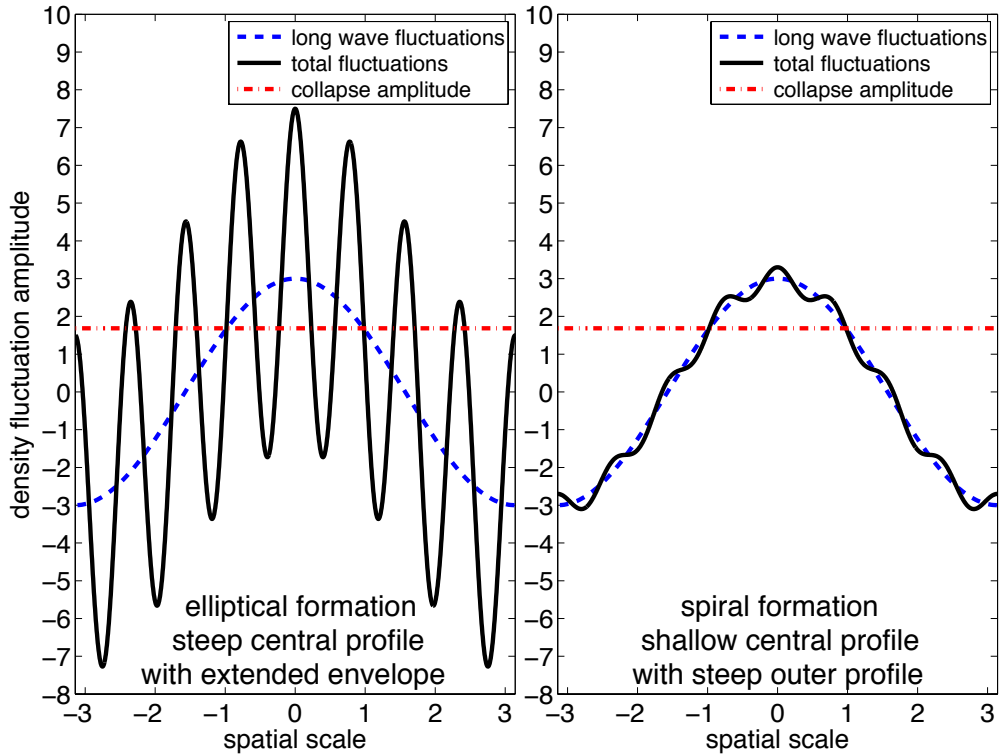}}   
\vskip -0.5cm
\caption{
{\color{burntorange}\bf Left panel:}
shows an example of the formation of a massive galaxy at $z=1$ that is overall determined by the large waves
indicated by the blue dashed curve. On top of the large linear wave, there is small-scale linear wave of length
$1/8$ of the large one with a fluctuation amplitude 1.5 times larger than the large wave.
Both long and short waves are chosen to be sinusoidal for the illustration.
The sum of the long and short waves is shown as the black solid curve.
The horizontal red dot-dashed line of amplitude value $1.68$ indicates the amplitude of the fluctuation that has collapsed by $z=1$.
{\color{burntorange}\bf Right panel:}
same as the left panel except 
the small-scale wave has a fluctuation amplitude 10 times smaller than the large wave.
Note that the examples are shown in 1-d but meant to be in 3-d.
}
\label{fig:sine}
\end{figure}

In a gaussian random field, different waves are superimposed on one another in a random fashion,
with the ensemble of waves at a given length following gaussian distribution and 
the square of the mean equal to the amplitude of the power spectrum at that wavelength.
Here, a simple illustration is shown to contain rich physics and can already account for the 
basic trend of the Sersic profiles, which, more importantly, are additionally in accord with properties 
of galaxies other than the profiles.

Figure~\ref{fig:sine} 
shows an example of the formation of a massive galaxy that contains small-scale fluctuations with large amplitude (left panel)
and an example of the formation of a massive galaxy that contains small-scale fluctuations with small amplitude (right panel).
In both panels peaks that are above the horizontal red dot-dashed line would have collapsed by $z=1$.
Our choice of redshift $z=1$ has no material consequence and we expect the generic trends
should not depend on that choice.

In the left panel we see that, between the two points where the blue dashed curve intersect the horizontal red dot-dashed line,
there are three separate density peaks with peak amplitude of $6-7$.
Thus, a significant portion of the three peaks would have collapsed by redshift $z=4-6$ to form
three separate galaxies. 
Note that structures formed at higher redshifts tend to be denser than structures formed at lower redshifts.
Therefore, these earlier structures would settle to form the dense central region.
Although it is probable that the galaxies formed at the three separate peaks
subsequently merge to form a dense elliptical galaxy, 
our conclusion of forming a dense central region in this case does not necessarily require all of them to merge.
Moreover, there are two somewhat smaller peaks at x values of $\sim -1.5$ and $\sim + 1.5$ of amplitude $\sim 4.5$,
which would have collapsed by redshift $z=2-4$.
In addition, there two still smaller peaks at x values of $\sim -2.5$ and $\sim + 2.5$ of amplitude $\sim 2.5$,
which would have collapsed by redshift $z=1-2$.
It is reasonable to expect that the four outer small galaxies would accrete onto the central galaxy to form the outer envelope by $z=0$.
Thus, this configuration would form a central dense structure with a steep profile due to the early formation of 
the central subunits and their subsequent descent to the center (and possible merging),
and an extended envelope due to later infall of small galaxies that form in outer regions at some earlier times, 
resulting in a profile resembling a Sersic profile with $n\gg 1$.
This overall picture seems to resemble two-phase formation scenario 
for elliptical galaxies from detailed cosmological hydrodynamic simulations \citep[][]{2010Oser, 2011Lackner}. 

In the right panel we see that, between the two points where the blue dashed curve intersect the horizontal red dot-dashed line,
there is no significant substructure.
Therefore, the collapse of the central region will be rather coherent without significant central condensation (i.e., without a stellar bulge).
Furthermore, there is no significant density peak outside the central region that has collapsed;
as a result, there is little stellar envelope due to late infall of small galaxies.
Thus, this configuration would form a galaxy with a shallow central density slope and a very steep outer slope.
We suggest that this configuration would form a bulge-less spiral galaxy with a profile similar to a Sersic profile with $n=1$.
A corollary is that the configuration depicted in the right panel would occur in a ``quiet" environment,
which may be quantitatively described as having a small pair-wise velocity dispersion 
\citep[][]{1983Davis} or a high Mach number \citep[][]{1992Suto}. 
Our local environment appears to belong to this category.
Perhaps this explains why there is preponderence of giant bulge-less galaxies in our neighborhood
\citep[][]{2010Kormendy}.
This does not necessarily suggest that the observed large fraction ($\sim 50\%$) of large bulge-less galaxies in our local universe is representative for the universe as a whole.
Our own expectation is that the fraction of large bulge-less galaxies, averaged over the entire universe,
will be substantially lower than that seen in the very local neighborhood.
Future surveys with resolutions as good as those for local galaxies now can check this.

It is easy to imagine a variety of configurations that may fall in-between these two (nearly) bookend examples.
Since the gaussian density fluctuation is ``compensated" in the sense
that the large density peak tends to be sandwiched by a pair of troughs,
the expected trend is this: a larger degree of central substructure is 
accompanied by a larger degree of substructure in the outskirts,
whereas a lesser degree of central substructure is accompanied by a lesser degree of substructure in the outskirts.
Since the total density fluctuations are linear combinations of each independent waves,
one can generalize the configurations from two waves 
to an arbitrary number of waves but the trend seen in Figure~\ref{fig:sine} remains.
In short, the generic trend obtained essentially hinges on two important features of the gaussian random field:
each density wave is {\it compensated} and {\it independent}.

\section{Discussion and Conclusions}

Based on a simple analysis we show that gaussian random field provides the physical origin for the observed Sersic profiles.
The two unique properties of the gaussian random field - waves are compensated and independent -
dictate that a more central concentrated stellar structure of a galaxy is simultaneously
accompanied by an extended stellar envelope, and vice versa. 
Additionally, those with steep inner slopes are expected to contain 
significant subunits that form early and coalescence later, which 
are consistent with the paradigm of merger driven formation of elliptical galaxies,
whether being dry \citep[][]{1982vanAlbada} or wet \citep[][]{2006Hopkins}. 
On the other hand, those with shallow inner slopes are expected to contain 
little substructure, which would bode well for the formation of disk galaxies.
Thus, the picture is self-consistent.

This analysis is illustrative and qualitative.
It will be useful later to formulate a model that is quantitative and statistical 
in the context of the gaussian field statistics \citep[e.g.,][]{1986Bardeen, 1991Bond}. 
The task in hand is, however, still more complex than a full statistical analysis of the gaussian random field,
because baryonic physics is expected to play an important role.
Cosmological reionization (a.k.a, photoheating of the intergalactic medium),
gravitational shock heating due to large-scale structure formation and feedback from
stellar evolution and supermassive black hole growth may quantitatively 
change the stellar makeup in both the ``central region" and ``outer region"
shown in Figure~\ref{fig:sine}, perhaps to varying degrees.
Nevertheless, we see no physical reason that any of these baryonic processes
will alter qualitatively the systematic trend that is illustrated in the previous section.

Since the examples shown in Figure~\ref{fig:sine} are generic in terms of spatial scales,
we expect that our argument is applicable on other scales. 
If the above analysis on galaxies is extended to clusters of galaxies,
the following new predictions are made.
Clusters of galaxies are expected to display a similar trend or a family of density profiles
from concentrated ones resembling those of elliptical galaxies with large Sersic index $n$ 
to less concentrated ones resembling those of disk galaxies with small Sersic index $n$.
For this purpose, the characterization of density profiles of clusters of galaxies 
should be performed with respect to the stellar component.
Procedurally, one first identifies the virial radius of the cluster and then finds
the best Sersic fit within the virial radius.
Two critical issues are that cluster members are properly identified
and projection effects minimized, and that intra-cluster light is accounted for.
As an example, clusters with cD galaxies should display the shallowest slope and the most extended 
distribution of galaxies in the outer regions,
resulting in a very high $n$ value if fit with Sersic profile for the stellar density.
This prediction is verifiable.
Although a tight correlation between the presence of cooling flows in X-ray clusters and the presence of cD galaxies 
at the center is observed, it becomes natural to expect such an outcome from our analysis,
because the presence of cD galaxies means early formation of at least the ``seed" of 
the central region that is denser in gas and dark matter as well as in stars.

\vskip 1cm

I thank an anonymous referee for an especially positive report and very useful suggestions
that have improved the paper.
This work is supported in part by NASA grant NNX11AI23G.


\end{document}